\documentclass[a4paper]{jpconf}
\usepackage{graphicx}
\begin{document}
\title{Partial wave analysis at BES~III harnessing the power of GPUs}

\author{Niklaus Berger}

\address{Institute of High Energy Physics, 19B Yuquan Lu, Shijingshan, Beijing 100049, China}

\ead{nberger@ihep.ac.cn}

\begin{abstract}
Partial wave analysis is a core tool in hadron spectroscopy. With the high statistics data available at facilities such as the Beijing Spectrometer III, this procedure becomes computationally very expensive. We have successfully implemented a framework for performing partial wave analysis on graphics processors. We discuss the implementation, the parallel computing frameworks employed and the performance achieved, with a focus on the recent transition to the \emph{OpenCL} framework.
\end{abstract}

\section{Introduction}

The Beijing Electron-Positron Collider (BEPC) II is currently the world's highest luminosity machine in the $\tau$-charm energy region. The Beijing Spectrometer (BES) III experiment was designed to make the best possible use of this luminosity \cite{Ablikim:2009vd}; one of the core parts of the physics program \cite{Asner2008} is a thorough study of the light hadron spectrum. For many final states, a wealth of, often broad, intermediate resonances will contribute to the amplitude. In order to disentangle interference effects and determine the spin and parity of the resonances, a partial wave analysis (PWA) is performed. As this involves the computation of the amplitude for every event in every iteration of a fit, this becomes computationally very expensive for large data samples. As events are independent and the amplitude calculation does not vary from event to event, this task is trivially parallelizable. This and the floating point intensity predestine PWA for implementation on graphics processing units (GPUs). We have successfully implemented a framework performing tensor calculations and partial wave fits on GPUs. Both the software architecture which completely encapsulates all the complexities of GPU based computing and the raw power of today's GPUs lead to significantly shortened analysis turnaround times.

\section{Partial wave analysis}

In a typical PWA (we use the radiative decay $J/\psi \longrightarrow \gamma X, X \longrightarrow K^+K^-$ \cite{Bai2003} as an example), the decay is modelled by coherently summing the contributions from a variety of intermediate resonances $X$. The relative magnitudes and phases of these resonances are determined from a fit and the fit result is compared to the data. The set of resonances and their properties are changed until a sufficient agreement with the data is found.
\newpage
The intensity $I$ (relative number of events) at a particular point $\Omega$ in phase space can be expressed as
\begin{equation}
	I(\Omega) = \left|\sum_{\alpha} V_{\alpha}A_{\alpha}(\Omega)  \right|^{2},
\end{equation}
where the sum runs over all intermediate resonances $\alpha$, $V_{\alpha}$ is the complex production amplitude for $\alpha$ and $A_{\alpha}(\Omega)$ the complex decay amplitude at a particular point in phase space. The likelihood for a particular model is
\begin{equation}
	\mathcal{L} \propto \prod_{i = 1}^{N_{Data}} \frac{I(\Omega_{i})}{\int \epsilon(\Omega) I(\Omega) d\Omega},
\end{equation}
where the product runs over all $N_{Data}$ events in the sample and the integral is proportional to the total cross section, corrected for the detector efficiency $\epsilon(\Omega)$. The integral is usually performed numerically by summing over a large number $N_{MC}^{Gen}$ of simulated events (Monte Carlo, MC) generated evenly in phase space. The limited acceptance and efficiency of the detector can be taken into account by summing only over the $N_{MC}^{Acc}$ simulated events that pass the reconstruction and analysis cuts. In an iterative fit, a minimum of the negative logarithm of the likelihood, corresponding to the best set of parameters for the used model is searched for;
\begin{equation}
	- \ln{\mathcal{L}} \propto - \sum_{i =1}^{N_{Data}} \ln\left( \sum_{\alpha} \sum_{\alpha\prime} V_{\alpha}V_{\alpha\prime}^{*} A_{\alpha}(\Omega_{i}) A_{\alpha\prime}^{*}(\Omega_{i})\right) \nonumber 	+  \sum_{\alpha} \sum_{\alpha\prime}  V_{\alpha}V_{\alpha\prime}^{*} \left( \frac{1}{N_{MC}^{Gen}} \sum_{j=1}^{N_{MC}^{Acc}} A_{\alpha}(\Omega_{j}) A_{\alpha\prime}^{*}(\Omega_{j})\right).
\end{equation}
The first sum runs over all data events, the second over all MC events. If the widths and masses of resonances are kept constant in the fit (i.e.~the $V_{\alpha}$'s are the only free parameters), the last (inner) bracket and the $A_{\alpha}(\Omega_{i}) A_{\alpha\prime}^{*}(\Omega_{i})$ term for each data event can be pre-calculated.

The number of floating point operations required is dominated by the sum over the data events and scales with $N_{iterations} \times N_{data} \times N_{waves}^2$, whilst the lookup table takes up storage space scaling with $N_{data} \times N_{waves}^2$. 
The storage space problem can be addressed by increasing the memory of the relevant machine ($\approx$1.5~GB are required per million events for a model with 20 partial waves), or with appropriate caching mechanisms for data samples with several million events. If the required floating point operations are performed sequentially, the time required can however become very long. There is no way of knowing whether the fit found just a local or the searched for global maximum of the likelihood. To gain confidence in the result, the fits are usually repeated with various sets of starting parameters. In addition, various models have to be tried out, especially in the study of possible new resonances and systematic effects. The thousands of fits needed in a typical partial wave analysis should thus be as fast as possible, especially as feedback from the physicist is required between the fits and they thus cannot easily be ran in parallel. 

\section{GPU computing}

The advent of computer games with realistic three-dimensional depictions of a virtual world computed in real time has pushed the development of ever more powerful graphics processing units (GPUs). GPUs provide a large number of floating point units optimized for executing the same code on many sets of input data (\emph{Single Instruction Multiple Data} SIMD). The theoretical performance of current GPUs surpasses 2~TFlop/s, at commodity prices of a few hundred dollars.

Whilst early efforts for using graphics processors for general purpose computing relied on graphics interfaces such as \emph{OpenGL} \cite{OpenGL3}, several frameworks for easy access to GPU computing have become available in the meantime. Besides the vendor supplied frameworks (\emph{CUDA} from Nvidia \cite{CUDA} and \emph{CAL/Brook+} from ATI \cite{AMDBrook}) there is now a vendor independent standard, \emph{OpenCL}~\cite{OpenCL, Munshi2010}. Most of these frameworks provide lots of low level control on the granularity of the parallelism, memory management and so on. For typical high energy physics applications (and in particular PWA), this fine-grained control is not really needed, as events are fully independent. For these cases, a framework with a high level of abstraction such as \emph{Brook+} is very convenient\footnote{We initially based our project on \emph{Brook+} and ATI hardware because at the time, only ATI allowed double precision calculations on consumer cards. This had the disadvantage that we were working with a rather experimental product with a small user base compared to Nvidias \emph{CUDA}. \emph{Brook+} however offers a higher level of abstraction, a narrow interface to the GPU and more elegant programming.}. As \emph{Brook+} is however not developed any further because AMD/ATI concentrates on \emph{OpenCL}, we built an abstraction layer on top of \emph{OpenCL}.  

\section{The GPUPWA framework}

We have developed a framework for GPU assisted partial wave analysis called \emph{GPUPWA}. It supports amplitude creation and calculation in the covariant tensor formalism on the GPU as well as GPU based likelihood calculations. Sophisticated caching mechanisms ensure that the memory footprint on the GPU stays within the rather tight limits (especially if using commodity hardware). The ROOT framework \cite{Brun1997} is used for data input and output, histogramming and, via the \emph{Minuit2} interface \cite{James1975}, for the minimisation process\footnote{For all cases where we can provide analytic gradients, we rely on the \emph{FUMILI} fitter \cite{FUMILI}, as it converges with the fewest iterations and we do not require reliable parameter errors from the fitter in most cases.}. 

The framework completely abstracts all GPU interna from the user, who can program in pure C++, allowing for the creation of a PWA within hours (few days in cases with many complicated amplitudes). More details on the interface and implementation (and also the striking similarities between PWA and game graphics calculations) can be found in \cite{Berger:2010zza}, which describes the \emph{Brook+} implementation, which is very close to the current \emph{OpenCL} version.

Most of the calculations on the GPU are performed with single precision floating point numbers. The final sum however is always performed in double precision. For a large enough number of events the final precision is comparable to a fully double precision CPU implementation or even better, as the parallel tree summing employed on the GPU has less rounding issues than a traditional accumulator sum.

\section{Performance}\label{sec:perf}

\begin{figure}[ht]
\begin{center}
\includegraphics[width=0.65\textwidth]{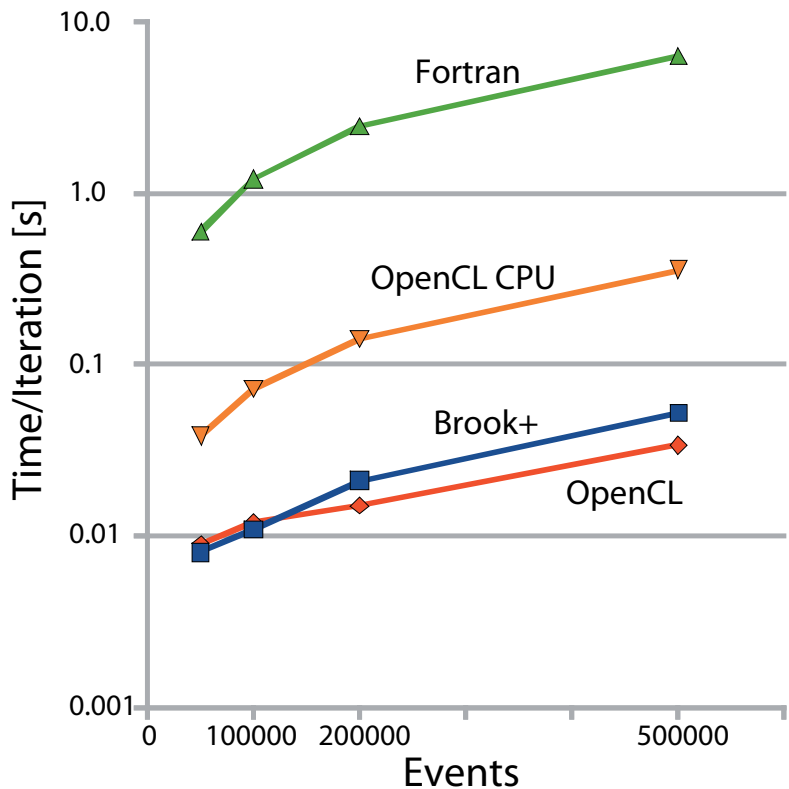}
\end{center}
\caption{\label{alltimes}Comparison of the running time of one fit iteration of the benchmark $J/\psi \longrightarrow \gamma K^+K^-$ analysis in different implementations on the same machine. Both \emph{OpenCL} datasets were obtained using the AMD implementation. The hardware used for the tests is described in section \ref{sec:perf}.}
\end{figure}

\begin{figure}[ht]
\begin{center}
\includegraphics[width=0.65\textwidth]{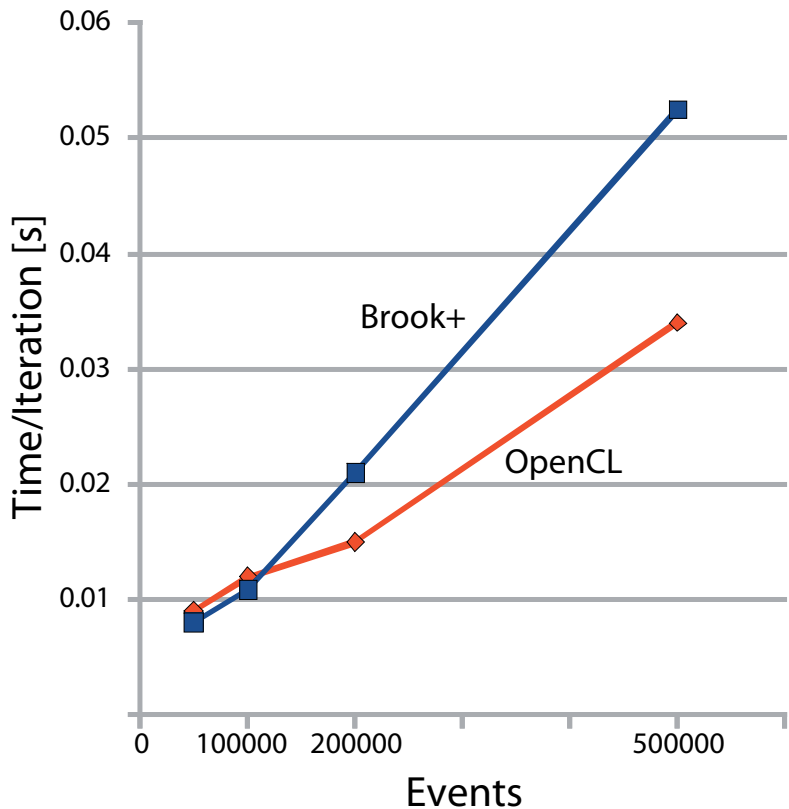}
\end{center}
\caption{\label{twotimes}Comparison of the running time of one fit iteration of the benchmark $J/\psi \longrightarrow \gamma K^+K^-$ analysis using the \emph{Brook+} and the \emph{OpenCL} framework (in the ATI implementation) on the same machine/GPU. Both \emph{OpenCL} datasets were obtained using the AMD implementation. The hardware used for the tests is described in section \ref{sec:perf}.}
\end{figure}

All the performance measurements quoted here were performed on an Intel Core 2 Quad 2.4~GHz workstation with 2~GB RAM running under Scientific Linux 5.2. The GPU is a ATI Radeon~4870 with 512~MB RAM\footnote{In the meantime, much more powerful hardware is available both at the CPU and the GPU end; the machine has however been available during the complete GPUPWA development process and thus remains an useful reference.}. We compared a FORTRAN implementation used in BES~II (without multi-threading and vectorization and fully in double precision) and the \emph{BROOK+} GPUPWA implementation of a $J/\psi \longrightarrow \gamma K^+K^-$ partial wave analysis, both minimizing using Fumili with analytical gradients and Hessian. The time for one fit iteration\footnote{Typical Fumili fits require 5-10 iterations, Minuit fits rather 50-100 (come however with useful error estimates).} was measured with the system timer for various sizes of the data samples (see figure \ref{alltimes}). During the fit, data transfers over the PCIe bus are restricted to the fit parameters (to the GPU) and likelihoods, gradients and Hessian matrix (to the CPU); bus bandwidth is thus not the limiting factor.
For large enough samples, the \emph{BROOK+} GPUPWA code is more than 150 times faster than the FORTRAN code. The same analysis was then ported to \emph{OpenCL} in the AMD/ATI implementation. Without further optimisation of the code, another 35\% speed-up was obtained for large event numbers (see figure \ref{twotimes}); in part this is due to fewer restrictions in the size of memory allocations in \emph{OpenCL} compared to \emph{Brook+}, the rest is probably due to a more efficient compiler.

The AMD/ATI \emph{OpenCL} implementation also allows for running the code purely on the CPU, making use of multiple cores and vector units. On our test machine, this leads to a speed-up of more than one order of magnitude with regards to the FORTRAN implementation, but also still another order of magnitude removed from the GPU performance.

\begin{table}
\caption{Time taken by an example analysis in various steps of the calculation in the \emph{OpenCL} implementation on the GPU. For details of the set-up, see section \ref{sec:perf}. Note that the \emph{fit} step includes initialization and cleanup and times are thus not directly comparable with those shown in figures \ref{alltimes} and \ref{twotimes}.\\}
\label{tab:times}
	\centering
		\begin{tabular}{lrr}
			\hline
			Step & Time [s] for   & Time [s] for  \\
					& 50'000 Events & 500'000 Events \\
			\hline
			Start-up (initialization and reading of files) & 0.75 & 3.67\\
			MC Integral           & 1.01 & 4.65\\
			Lookup table creation & 1.64 & 4.92\\
			Fit (5 iterations)    & 0.10 & 0.99\\
			Plots                 & 1.48 & 9.82\\
			\hline
			Total                 & 4.98 & 24.05\\ 
			\hline
		\end{tabular}
\end{table}

A rough estimate on the number of floating point operations (both in single and double precision) performed indicates that we reach on the order of 10~GFlop/s on the GPU, about 1~\% of the theoretical floating point performance of the card; most of the time is thus spent in memory accesses, communication over the PCI bus and the \emph{MINUIT}/Fumili step on the CPU. As however the run time of a typical fit is dominated by reading input data and producing plots (see table \ref{tab:times} for example total timings), we do not put a high priority on further optimization of the fit proper. 

\section{Open issues}

\emph{OpenCL} finally offers a vendor independent access to these resources. \emph{OpenCL} leaves however many details of the implementation to the vendors, possibly leading to portability issues; our attempt to port the \emph{OpenCL} version of GPUPWA to an \emph{Apple} computer running the \emph{Snow Leopard} operating system (which natively incorporates \emph{OpenCL}), was hampered by the different implementation of vector data types in the CPU part of the code between Apple and AMD (GPUPWA uses the \emph{OpenCL} vector data types also in the C++ part of the code, e.g.~for initialisation of data structures and event-independent tensor manipulations). Obviously it would be desirable to overcome these portability issues in the future. It would also be helpful for many high energy physics applications if there would be a higher level interface to the GPU than \emph{OpenCL}, something like \emph{Brook} for \emph{OpenCL}.
\newpage

Most of the technical issues in PWA\footnote{There is also a wealth of issues related to the physics, especially as to what amplitude models to employ. These issues at the interface between theory, experiment and computing have been discussed in depth in recent workshops in Seattle \cite{Seattle} and Trento \cite{Trento}.} are however related to the fitting; in typical fits there are 20 to 100 free parameters, with all the issues that brings with it. Most of the parameters represent complex numbers, which are not handled well by the usual fitters; if a Cartesian representation is used, large correlations between the components are created if the phase is not well constrained, hampering fit convergence. A polar representation relies on a bounded and a periodical parameter, the first of which problematic for the fitter if the bound is a valid value, the second not handled by the fitters at all and often requiring manual intervention. A minimizing algorithm for complex numbers (including the proper handling of the derivatives) would be extremely useful here. 

\section{Conclusion and outlook}

Using massively parallel computing, especially in the for of GPUs, is a powerful tool to overcome speed limits in partial wave analyses. The GPUPWA framework developed at the BES~III experiment offers easy access to these resources with a pure C++ interface and all core functionalities provided by the framework. GPUPWA has been ported to use the \emph{OpenCL} standard to interface parallel computing devices. The speed gain compared to a single-threaded non-vectorized version is more than an order of magnitude if using multiple threads and vector units on the CPU and more than two orders of magnitude when employing the GPU. Future development will focus on adding physics functionality to GPUPWA and to make the application truly portable. This will then also allow for hardware comparisons between the major GPU vendors.  

\ack

I would like to thank my colleagues at IHEP who are testing and using GPUPWA for their many helpful suggestions, bug reports and additions to the framework.

The work presented was supported by the Chinese Academy of Sciences and the Swiss National Science Foundation.

\section*{References}
\bibliography{Taipeh}

\end{document}